# Comparison of experimental and simulation distortions of quenched C-ring test parts

Nicolas Cyril[1*], Baudouin Cyrille[1], Leleu Stéphane[2], Teodorescu Mihaela[3] and Bigot Régis[1]

[1] Arts et Metiers ParisTech, LCFC, 4 rue Augustin Fresnel, 57070 Metz – France
[2] Ascometal Research Center (CREAS), BP 70045 – 57301 Hagondange – France
[3] Arts et Metiers ParisTech, L2MA, 8 boulevard Louis XIV, 59046 Lille – France

**ABSTRACT:** We present a mathematical method for identifying, separating and quantifying the 3D significant distortions. Measurement of a gas quenched C-ring type sample is performed by a Coordinate Measuring Machine (CMM). Quenching simulation is done with the commercial software Forge 2008 TTT®. After comparisons, we notice that the distortions and their tendencies are the same but not exactly the amplitudes. We only focus on distortions which have a physical origin, like the pincers opening, often observed in the literature. As a first explanation, we underline the role of phase transformations and volume dilatation of the steel during quenching.

## 1 INTRODUCTION

Quenching is valuable to improve product mechanical properties but may increase the product cost by inducing undesirable dimensional changes. Nowadays, one of the main industrial goals is to minimize these changes. To do so, modelling is an interesting tool to predict distortions. However, distortions prediction is complex, its accuracy depending on many parameters (thermal, mechanical and metallurgical) which are, mostly, experimentally estimated.

In a first part, we introduce the optimization method used in this study. The second part deals with the experimental approach for accurately identifying distortions. Then we give the simulation conditions used to predict distortions. Finally, a comparison between a measured and simulated C-ring is provided, for an ASCOMETAL steel grade. We also propose explanations of the origins of the distortions phenomena.

## 2 DEVELOPED METHOD

### 2.1 Elementary defects dissociation

Figure 1 illustrates the three steps of our analysis method. The first step (❶) is the measurement of the discrete geometry of the part before and after heat treatment. For that, we use a Coordinate Measuring Machine (CMM) because it is a flexible machine commonly used in industry. The second step (❷) is the relative comparison between measurement points before and after quenching. The goal is the decomposition of the overall distortion in elementary phenomena, significant on a macroscopic scale and having a physical reality. We thus qualify their geometrical signatures, i.e. their single prints and we classify them in two main groups:
- the positioning defects (translations and rotations);
- the form defects (radius dilatation, "bobbin" effect).

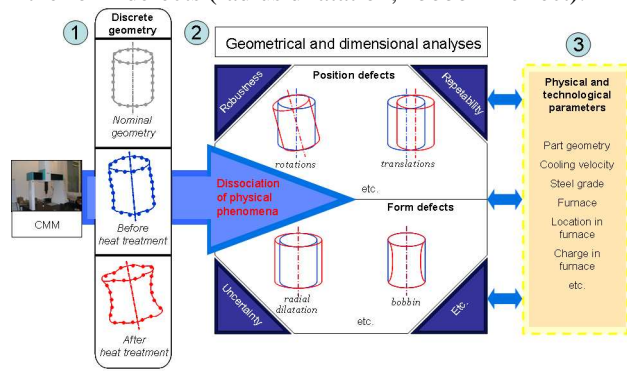

**Figure 1:** *Our method for identifying distortions*

Distortions are linked to physical and technological parameters: phase transformation, residual stress, steel grade (step ❸). Hence, the thinking on the physics of the phenomena brings us to the writing of their signature and thus their quantification, with an uncertainty value. This uncertainty is related to the systematic and random errors of the measurement process.

### 2.2 Need for using an optimization method

Mostly, the number of measurement points is higher than the number of distortions phenomena. The obtained system is thus redundant and does not have any analytical solution. Its solving requires an optimization method. The main approaches are geometrical [1], modal [2] [3] and genetic [4]. Since some methods use generic decomposition models, it is not easy to explain the physical origin of some elementary defects. That is why

---

* Nicolas Cyril: 4, rue Augustin Fresnel – 57078 Metz Cedex3, 03 87 37 54 30, cyril.nicolas@metz.ensam.fr

we developed a method optimizing the "signatures vectors" of the main physical phenomena [5].

### 2.3 Mathematical approach
#### 2.3.1 Data input of the physical process
The solution of the data-overabundant system is given by the orthogonal projection of the measurement vector $\varepsilon$ onto an analysis basis made of the $p$ identified defects, called $M_{ph}$. In Figure 2, we consider two phenomena vectors ($Ph_1$ and $Ph_2$) forming a plane (P) as the analysis basis. The residual vector corresponds to the measurement field which cannot be explained by only these two phenomena. The quantity thus minimized is the Euclidian norm of the residual vector, which corresponds to the least squares criterion, commonly used. Indeed, it is stable for an initial solution closed enough to the final one and it gives a unique solution.

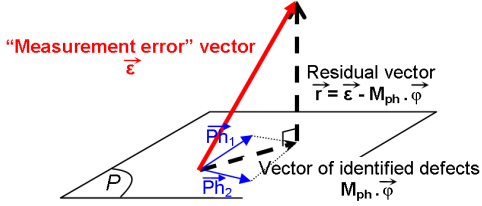

**Figure 2:** *Definition of the "analysis basis": $M_{ph}$*

#### 2.3.2 Fundamental hypothesis
The dissociation of distortion phenomena is possible when the two following assumptions are verified:
- Linearity: firstly, the mathematical function of phenomena must be linear. Secondly, they must be linearly superimposable, in order to be dissociated;
- Independence: phenomena must be linearly independent. If a complete dependence leads to the impossibility to resolve the system, the problem becomes more delicate when phenomena have closed effects without however being identical. The propagation of the measurement uncertainty is used to judge the accuracy of the distortions vectors of the analysis basis.

#### 2.3.3 Steps of the solving
Equation (1) provides the measurements vector $\varepsilon$.

$$\vec{\varepsilon} = \begin{pmatrix} \varepsilon_1 \\ \vdots \\ \varepsilon_i \\ \vdots \\ \varepsilon_n \end{pmatrix} \quad \left\| \begin{array}{l} \varepsilon_i = (\overrightarrow{OT_i} - \overrightarrow{OM_i}).\vec{N_i} \\ \|\vec{N_i}\| = 1 \end{array} \right. \quad (1)$$

Its scalars correspond to the errors between theoretical points $T_i$ and measured points $M_i$, orthogonally projected onto theoretical normals $N_i$. Then, we define in Equation (2), the matrix $M_{ph}$ for the p signatures. The scalars correspond, for an unit amplitude, to the phenomena effect ($e_{(Ph_i)_j}$) on normals, for each measurement points.

$$M_{ph} = \begin{Bmatrix} & \overrightarrow{Ph_1} & \cdots & \overrightarrow{Ph_j} & \cdots & \overrightarrow{Ph_p} \\ \vec{\varepsilon_1} & e_{(Ph_1)_1} & \cdots & e_{(Ph_j)_1} & \cdots & e_{(Ph_p)_1} \\ \vdots & \vdots & \ddots & \vdots & \ddots & \vdots \\ \vec{\varepsilon_i} & e_{(Ph_1)_i} & \cdots & e_{(Ph_j)_i} & \cdots & e_{(Ph_p)_i} \\ \vdots & \vdots & \ddots & \vdots & \ddots & \vdots \\ \vec{\varepsilon_n} & e_{(Ph_1)_n} & \cdots & e_{(Ph_j)_n} & \cdots & e_{(Ph_p)_n} \end{Bmatrix} \quad \begin{array}{l} \text{with, for example, } \overrightarrow{Ph_1} = \overrightarrow{Tx} \\ \text{and } \|\overrightarrow{Tx}\| = 1 \\ \Rightarrow e_{(Ph_1)_i} = \overrightarrow{Tx}.\vec{N_i} = \vec{N_i}.\vec{x} \end{array} \quad (2)$$

The real solution of the system $M_{ph}.\varphi = \varepsilon$ is given by $\varphi = (M_{ph}^t.M_{ph})^{-1}.M_{ph}^t.\varepsilon$. Its solving with an optimization in the sense of least squares is equivalent to minimize the norm of the residual vector (Equation (3)). We finally get the $a_p$ scalars of the vector $\varphi$, which are the proportional amplitudes of each one of the $p$ defects.

$$\begin{array}{c} M_{ph}.\varphi = \varepsilon \Leftrightarrow \min_{\phi \in \mathfrak{R}^p} \|\vec{\varepsilon} - M_{ph}.\vec{\varphi}\|^2 \Leftrightarrow \min \|\vec{r}\|^2 \\ \text{with} \\ r = \left\| \vec{\varepsilon} - a_1.\overrightarrow{Ph_1} - a_2.\overrightarrow{Ph_2} - a_3.\overrightarrow{Ph_3} - \ldots - a_p.\overrightarrow{Ph_p} \right\| \end{array} \quad (3)$$

### 2.4 Propagation of the measurement uncertainty
It is an indicator to describe the pertinence of the analysis basis. Indeed, the increase of phenomena improves the risk of slight dependence, which disturbs the solving. Since measurements are done with an uncertainty, called $\sigma_i$, the $a_p$ amplitudes of the phenomena are consequently affected by an uncertainty, called $\sigma_{a_p}$. We consider a probabilistic approach [6]: each $\varepsilon_i$ is the realization of a random variable $E_i$, which obey a normal law N ($\beta_i$, $\sigma_i^2$). If we replace all $\varepsilon_i$ with their expectations $\beta_i$, the system $\mathbf{M_{ph}.\varphi = \varepsilon}$ becomes $\mathbf{M_{ph}.u = \beta}$, $u \in R^n$. After mathematical proof, we obtain the expression of $\sigma_{a_p}$ in Equation (4).

$$\sigma_{a_p} = \sqrt{a_{ii}} \text{ with } a_{ii} \text{ the } i^{th} \text{ diagonal elements of matrix} \left( \frac{M_{ph}^t.M_{ph}}{\sigma_i} \right)^{-1} \quad (4)$$

The value of $\sigma_i$ is obtained by measuring on a CMM, a gauge ring. The same experimental conditions as for the C-rings were used to take into account systematic and random errors of the measurement process. The obtained standard deviation $\sigma_i$ of all $\varepsilon_i$ is equals to 1.6 µm, for a confidence level of 99.7% (k=3).

## 3 EXPERIMENTAL APPROACH
### 3.1 Introducing C-ring test part
The C-ring (Figure 3) is used because its geometry highlights distortion [7]. Its dimensions minimize side effects and allow obtaining required cooling velocity. Concerning the measurement strategy, we use a fine mesh to accurately identify unknown distortions. Points are as best as possible, evenly distributed on all surfaces. We use a dynamic probe (TP20) with a star stylus and an extension bar. Combined with head rotations of the CMM probe, we reached all surfaces.

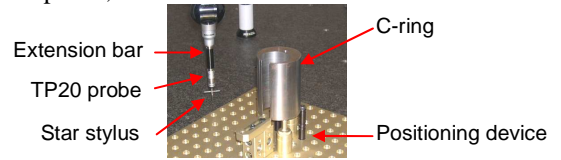

**Figure 3:** *C-ring test part and its measuring device*

## 3.2 Modelling distortions phenomena of a C-ring

We make now a relative comparison between the discrete geometry before and after gas quenching, with a steel grade 1 (Figure 4). For a better visual understanding, a scale factor of 50 is applied on the errors onto the theoretical normals. After having modelled distortions, we subtract their effects and we re-analyse the remaining residual errors. We repeat this iteration until the residuals are small enough.

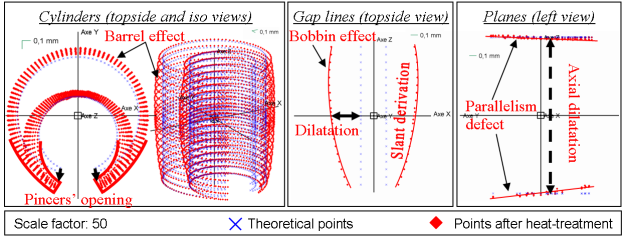

**Figure 4:** *C-ring distortions after heat treatment*

### 3.2.1 For cylinders

We clearly notice an opening of pincers. We model it by physically taking into account the opening of the neutral fibre of pincers, i.e. the line formed with the centres of the cylinders both tangential to the outer and inner cylinders. Mathematically, the pincers opening is modelled by a half period of $\cos\alpha$ function. We also detect a barrel phenomenon. It is characterized by a maximum amplitude at the middle of each generatrix and by no variation at the top and bottom. Mathematically, the barrel effect is modelled by a square function, optimized for each of the two cylinders.

### 3.2.2 For the planes and the gap

We take into account an axial dilatation of planes and a parallelism defect. These distortions are caused by the heating, generating stress areas on the surface of the outer cylinder and compression zones on the inside cylinder. Concerning the two generatrixes of the gap, we detect an x-axis dilatation, a bobbin effect (the opposite effect of barrel) and a slant derivation. These defects are obviously related with those of cylinders.

## 3.3 Expression of each phenomena's signature

In Figure 5, we present the 3D-mathematical expression for distortions, added in the analysis basis $M_{ph}$. Their relevance is judged by their efficiency to minimize the residual and their physical reality. We also draw their effect onto theoretical normals, with a scale factor of 5.

| Phenomena | Mathematical expressions | Effects on errors projected onto theoretical normals |
|---|---|---|
| Pincers opening | For each section of j=82 points: $\begin{cases} e_{open_j} = 1 - \cos\dfrac{\alpha_j}{2} \\ \alpha_j = \text{Atan2}(\overrightarrow{OT_j}.\vec{x}; \overrightarrow{OT_j}.\vec{y}) \end{cases}$ | |
| Barrel effect | For each generatrix of j=21 points: $\begin{cases} e_{barrel_j} = 1 - z_j^2 \\ z_j = -1 + 0.1 \times k, k \in [0,20] \end{cases}$ | |
| Dilatation of planes | For each plane: $\begin{cases} e_{dila\_plane_j} = \overrightarrow{N_j}.\vec{z} = Nz_j \\ \|Nz_j\| = 1 \Rightarrow e_{dila\_plane_j} = 1 \end{cases}$ | |
| Parallelism defect of planes | For each plane: $\begin{cases} e_{parallelism_j} = T_jz.Ny_j - T_jy.Nz_j \\ T_j \text{ is the jth theoretical point} \end{cases}$ | |

**Figure 5:** *Linear and independent distortion signatures*

## 4 SIMULATION APPROACH

### 4.1 Heating and quenching the part

We use Forge 2008 TTT software for 3D simulations. The input data are partly experimentally determined (phase transformation kinetics, heat transfer coefficients) and partly from the literature (mechanical data) [5].
C-ring is heated until the temperature reaches the same as for experiment (930°C) and is uniform in the part. The thermal, mechanical and metallurgical parameters are used, especially the phase transformations from the initial structure (ferrito-perlitic) at 20°C to the final austenitic structure at 930°C. In Figure 6 a), we clearly notice distortions and a great dilatation of the nominal geometry, with a maximum value of 1.6 mm for the outer cylinder and 1 mm for the inner.
Then, we used this distorted heated geometry to compute the gas quenching simulation. We didn't consider a cooling time, because the furnace carries quickly and automatically the heated C-rings to the vacuum gas chamber. So, after quenching, geometry is closer to nominal one (Figure 6 b)) but we note the same distortions observed in experiments: pincers opening, barrel effect, parallelism defect of planes. Like distortions phenomena can not be dissociated with the simulation software, we apply our optimization method.

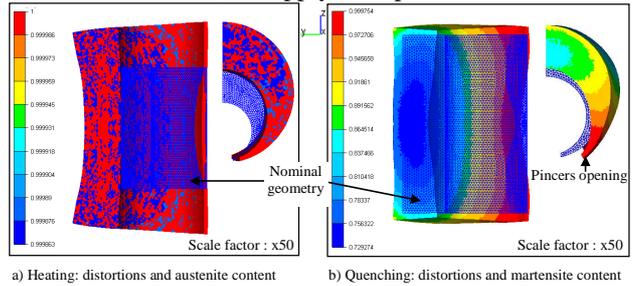

a) Heating: distortions and austenite content    b) Quenching: distortions and martensite content

**Figure 6:** *Distortions of an half simulated C-ring*

### 4.2 Data processing strategy

For a direct comparison between experimental and simulation results, we developed an algorithm which virtually measures the simulated part, as the CMM does. Indeed, a measurement point is obtained by probing the surface following a theoretical normal. So, we have to find the intersection point $M_1$ between theoretical normal $N_1$ and finite element plane $(S_1S_2S_3)$, formed with the three closest simulated points to theoretical point $T_1$. We thus get (Figure 7) the measurement vector $\varepsilon$. This virtual measurement induces an uncertainty, which is minimized by taking a great number of points for the mesh.

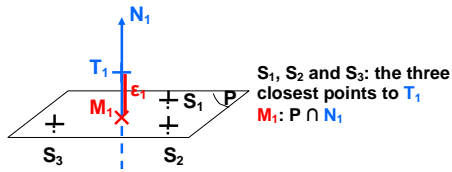

*Figure 7: Virtual measurement with simulated points*

## 5 COMPARISON OF RESULTS

### 5.1 Quantitative evaluation

The optimization is done with all distortions phenomena previously identified on cylinders, planes and gap. However, we focus on cylinders because they represent the main distortions in regards with their great number of points. So, we give in Figure 8, their $a_p$ values of distortions with their uncertainties $\sigma_{a_p}$. Firstly, the chosen vectors explain the main distortions as the final residual is small. Secondly, there are no dependent phenomena because uncertainties are small compared with $\sigma_i$ (1.6 μm) and $a_p$ values.

We detect the pincers opening both in experiment and simulation, as noticed in other study on C-rings [7] [8]. This distortion is also the most important. Values for simulation and experiment are different but tendencies are the same. The barrel effect is greater for the inner cylinder than for the outer.

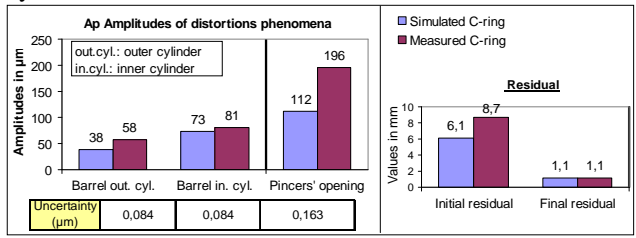

*Figure 8: Amplitudes of distortions and residual*

### 5.2 Physical origins

The pincers opening is a time-dependent combination between the thickness gradient, the velocity cooling gradient (faster near pincers edges) and steel phase transformations. The thermal gradient between the inner and outer cylinders plays an important role. Indeed, with a null gradient, they are closed up to a maximal value but the increase of the gradient makes bigger the opening.

Concerning the barrel effect, it could be compared to a shaft bending, as studied in [9]. So, it probably comes from both the heterogeneity of the cooling fluid velocity inside the inner and outer cylinder and the geometry. Consequently, it can be minimized by reducing the drasticity and thus, the pressure of gas.

## 6 CONCLUSIONS

By considering all points taken on the geometry of a quenched C-ring, we use an optimization method for identifying, quantifying and separating significant distortion phenomena. Distortions are modelled by taking into account their physical origin. Distortions amplitudes are quite the same in experiment and simulation but we have to refine input data of the models for a better quantitative evaluation. Moreover, the explanation of the physical origins of distortions requires their separation by carrying out a parametric analysis, by subtracting, for instance the thermal influence. Further works will also take into account the volume change by computing the pincers thickness variations.

## ACKNOWLEDGEMENT
The authors would like to express their special thanks to the ASCOMETAL CREAS Company for their scientific, technical and financial supports and the Province of Lorraine for its financial support.

## REFERENCES

[1] G.L. Samuel, M.S. Shunmugam: Evaluation of Circularity from Coordinate and Form Data Using Computational Geometric Techniques. *Precision Engineering*, 24: 251-263, 2000.

[2] W. Huang, D. Ceglarek: Mode-based Decomposition of Part Form Error by Discrete-Cosine-Transform with Implementation to Assembly and Stamping System with Compliant Parts. *Annals of the CIRP*, 51: 21-26, 2002.

[3] S. Samper, F. Formosa: Form Defects Tolerancing by Natural Modes Analysis. *Jour. of Computing and Information Science in Engineering*, 7: 44-51, 2007.

[4] T. Killmaier, A.R. Babu: Genetic approach for automatic detection of form deviations of geometrical features for effective measurement strategy. *Precision Engineering*, 27: 370-381, 2003.

[5] C. Nicolas, C. Baudouin, S. Leleu, M. Teodorescu, R. Bigot: Dimensional control strategy and products distortions identification. *Int. Jour. of Material Forming*, 2008.

[6] G. Diolez. Maîtrise de la position géométrique des solides : vers de nouveaux outils plus efficaces, *PhD thesis in french*, Arts et Métiers ParisTech, Lille, 2006.

[7] R.A. Hardin, C. Beckermann: Simulation of Heat Treatment Distortion. In *59th Technical and Operating Conference*, Chicago, 2005.

[8] Z. Li, B.L. Ferguson, X. Sun, P. Bauerle: Experiment and Simulation of Heat Treatment Results of C-Ring Test Specimen. In *23rd ASM Heat Treating Society Conference*, Pittsburgh, pages 245-252, 2005.

[9] M.Teodorescu, J. Demurger, J. Wendenbaum: Comprehension of Distortion Mechanisms of Gas Quenched Automotive Shafts. *Proceedings of 17th IFHTSE Congress*, Kobe, Japan, 2008.